\documentclass[aps,prd,reprint]{revtex4-2}

\usepackage{amsmath,amssymb,physics}
\usepackage{graphicx}
\usepackage{caption}
\usepackage{verbatim}

\usepackage[colorlinks=true,
            linkcolor=blue,
            citecolor=blue,
            urlcolor=blue]{hyperref}

\begin{document}

\title{Charge-Odd Hyperon Polarization from Magnetic Spin Precession}

\author{Dushmanta Sahu}
\email{Dushmanta.Sahu@cern.ch}
\affiliation{Instituto de Ciencias Nucleares, UNAM, Apartado Postal 70-543, Coyoacán, 04510, México City, México}

\author{Captain R. Singh}
\email{captainriturajsingh@gmail.com}
\affiliation{School of Physical Sciences, National Institute of Science Education and Research,
Jatni, Odisha–752050, India}

\begin{abstract}
We demonstrate that polarized strange quarks undergo Larmor precession in the intense magnetic field produced in non-central heavy-ion collisions and as a possible source of charge-odd hyperon polarization. Strange and antistrange quarks carry opposite electric charges and therefore acquire opposite precession phases. This opposite spin rotation mixes the transverse and longitudinal polarization components, yielding measurable polarization splittings between $\Lambda$ and $\bar{\Lambda}$ hyperons. For the magnetic-field evolution scenarios, the predicted splittings reach the sub-percent level and are within the experimentally accessible range at RHIC and the LHC energies. These charge-resolved hyperon polarization observables provide a direct probe of magnetic-field-driven spin dynamics of deconfined QCD matter at ultra-relativistic heavy-ion collisions.

%We show that polarized strange quarks undergo Larmor precession in the magnetic field generated in non-central heavy-ion collisions. Because strange and antistrange quarks carry opposite electric charges, they acquire opposite precession phases, producing charge-odd polarization splittings between $\Lambda$ and $\bar{\Lambda}$ hyperons through the mixing of transverse and longitudinal polarization components. For different magnetic-field evolution scenarios, the resulting splittings reach the sub-percent level and are experimentally accessible at RHIC and the LHC. These charge-resolved polarization observables establish hyperon polarization as a direct probe of magnetic-field-induced spin dynamics and the time-integrated magnetic field of the quark--gluon plasma.
\end{abstract}

\maketitle

\section{Introduction}

The possible existence of  quark--gluon plasma (QGP) in relativistic heavy-ion collisions is subjected to the strongest electromagnetic fields known in nature. In non-central collisions, fast-moving spectator protons generate magnetic fields in the order of $eB \sim m_\pi^2$ at the early stages of the collisions~\cite{Tuchin:2013apa,McLerran:2013hla,Basar:2012bp}. While the initial field strength is relatively well understood, its subsequent evolution remains one of the major open questions in heavy-ion physics. In the vacuum, magnetic field decays rapidly as the spectators separate. However, electromagnetic currents induced in the conducting QGP may significantly prolong its lifetime~\cite{Inghirami:2016iru,Huang:2022qdn}. Therefore, investigating the temporal evolution of the magnetic field is essential for understanding a broad range of electromagnetic phenomena in deconfined QCD matter. Recently, the interplay between rotation and magnetic field has been investigated in terms of Barnett and Einstein-de Haas effects~\cite{Sahu:2025tmb,Sahu:2026lbv,Sahu:2026ixs}.

On the other hand, hyperon polarization measurements have emerged as powerful probes for studying the rotational dynamics of the QGP medium~\cite{STAR:2017ckg,Becattini:2013vja,Sahoo:2024egx,Sahoo:2024yud}. The observation of global $\Lambda$ and $\bar{\Lambda}$ polarization established that the large orbital angular momentum generated in heavy-ion collisions can be transferred to particle spin~\cite{STAR:2017ckg,STAR:2018gyt}. Since then, extensive theoretical efforts have focused on understanding the generation of polarization through thermal vorticity, spin hydrodynamics, and kinetic theory approaches~\cite{Becattini:2017gcx,Florkowski:2017ruc,Weickgenannt:2019dks,Liu:2021uhn}. However, recent developments in relativistic spin hydrodynamics and spin magnetohydrodynamics open the possibility of investigating dynamical evolution of spin polarization under strong electromagnetic fields~\cite{Liang:2004ph,Fang:2016vpj,Bhadury:2022ulr}. A polarized quark carrying a finite magnetic moment can undergo Larmor precession in an external magnetic field, implying that the polarization vector may change its direction while medium is evolving~\cite{Jackson}. Consequently, the spin polarization observed at QGP phase boundary need not coincide with that initially generated in the medium, rendering polarization measurments sensitive to the space--time evolution of the electromagnetic field. 

Global hyperon polarization measurements probe spin alignment along the angular-momentum direction of the medium and provide indirect evidence of the interplay between vorticity and electromagnetic fields in the QGP. However, these fail to retain detailed information about the time evolution of the electromagnetic field. In contrast, transverse and longitudinal polarization components, such as $P_{x}$ and $P_{z}$, retain sensitivity to the dynamical spin evolution induced by Larmor precession in a time-dependent long-lived magnetic field. In this framework, the polarization vector evolves through the accumulated spin-precession phase, which encodes the dynamics of the magnetic field experienced by the strange quark prior to hadronization. As strange and antistrange quarks possess opposite electric charges, they acquire opposite precession phases in the presence of the magnetic field. Consequently, the polarization vectors of $\Lambda$ and $\bar{\Lambda}$ hyperons undergo opposite rotations, generating a charge-odd mixing between transverse and longitudinal polarization. This produces a particle--antiparticle asymmetry that cancels in charge-averaged samples but survives in charge-resolved measurements. Unlike global polarization, this mixing arises from electromagnetic spin evolution and provides a distinct signature compared with conventional vorticity-induced polarization mechanisms. Observing this signal requires charge-resolved measurements that analyze $\Lambda$ and $\bar{\Lambda}$ separately, as the charge-odd differences cancel entirely in the combined $\Lambda+\bar{\Lambda}$ measurements~\cite{ALICE:2021pzu,STAR:2019erd}. Thus, while existing measurements have focused on combined samples, future high-statistics runs with charge resolution can directly probe the in-medium dynamics of the magnetic field.\\

In this letter, we demonstrate that Larmor precession in a time-dependent magnetic field induces a characteristic charge-odd mixing between transverse and longitudinal hyperon polarization components. We show that the resulting $\Lambda$--$\bar{\Lambda}$ splittings are governed by the accumulated spin-precession phase, establishing hyperon polarization as a probe of the time-integrated magnetic field of the quark--gluon plasma. This mechanism provides a new experimental avenue to constrain the time-integrated magnetic field in relativistic heavy-ion collisions through charge-resolved polarization measurements.

\section{Formalism}

The strong magnetic field produced in non-central heavy-ion collisions can influence the evolution of spin-polarized quarks traversing through the QGP medium~\cite{Kharzeev:2007jp,Hattori:2016emy}. While most studies of hyperon polarization focus on the generation of spin polarization due to thermal vorticity and related phenomena~\cite{Becattini:2013vja,Becattini:2022zvf,Becattini:2020ngo}, an equally important question concerns the subsequent evolution of an inherently polarized system in the presence of a magnetic field~\cite{Bargmann:1959gz}. Since spin precession is a consequence of the interaction between a magnetic moment and an external magnetic field, it is largely independent of the microscopic origin of the initial polarization.\\

We consider polarized strange quarks propagating through a time-dependent magnetic field. In the local rest frame of the fluid, where $\mathbf{B}\simeq B_y\hat{y}$ dominates the electromagnetic field, the Bargmann--Michel--Telegdi (BMT) effect~\cite{Bargmann:1959gz} reduces to Larmor precession when we neglect electric field effects and relativistic corrections. These corrections affect only the accumulated precession phase $\Theta$. The charge-odd polarization splitting is instead a robust consequence of the opposite electric charges of strange and antistrange quarks. The polarization vector $\mathbf{P}$ then evolves as;

\begin{equation}
\frac{d\mathbf{P}}{d\tau} = \boldsymbol{\Omega}_{L}(\tau)\times \mathbf{P},
\label{eq1}
\end{equation}

where $\boldsymbol{\Omega}_{L}$ denotes the instantaneous Larmor frequency and $\tau$ is the proper time. The Larmor frequency is defined as;

\begin{equation}
\boldsymbol{\Omega}_{L} (\tau) = \frac{g_s q_s e}{2m_s}\mathbf{B},
\label{eq2}
\end{equation}

here $g_s\simeq 2$ is the quark gyromagnetic factor and $e$ is the electric charge and $m_s$ is the effective mass. For a strange quark of electric charge $q_s=-1/3$. In relativistic heavy-ion collisions, the dominant magnetic field is approximately aligned with the global angular momentum of the collision system, typically chosen as the $y-$axis, $\mathbf{B} = B_y(\tau)\hat{y}$. Thus, Eq.~\ref{eq1} reduces to,

\begin{align}
\frac{dP_{x}}{d\tau} = -\Omega_L(\tau)P_{z},\nonumber
\\
\frac{dP_{y}}{d\tau} = 0,\nonumber
\\
\frac{dP_{z}}{d\tau} = \Omega_L(\tau)P_{x},
\label{eq3}
\end{align}

where,

\begin{equation}
\Omega_L(\tau) = \frac{|q_s|eB_y(\tau)}
{m_s(\tau)}.\nonumber
\end{equation}
For strange quarks ($q_s = -1/3$), Eq.~(\ref{eq3}) describes the precession dynamics. Antistrange quarks ($q_{\bar{s}} = 1/3$) precess in the opposite direction due to the opposite electric charge, corresponding to the replacement $\Omega_L \rightarrow -\Omega_L$. We see that the polarization component parallel to the magnetic field remains unchanged, while the transverse and longitudinal components mix through spin precession. The evolution is completely determined by the accumulated spin-precession phase,

\begin{equation}
\Theta = \int_{\tau_0}^{\tau_f}
\Omega_L(\tau) d\tau,\nonumber
\end{equation}

which can be written explicitly as,

\begin{equation}
\Theta = \int_{\tau_0}^{\tau_f}
\frac{|q_s|eB_y(\tau)}
{m_s(\tau)}d\tau.
\label{eq4}
\end{equation}

For numerical estimates, we employ a constant effective strange-quark mass $m_s=0.45 \pm 0.1$ GeV, representative of typical strange-quark quasiparticle masses used in phenomenological descriptions of the QGP ~\cite{Peshier:1995ty,Bluhm:2007cp}. In principle, the effective mass may depend on temperature and proper time. Such variations primarily rescale the accumulated precession phase through $\Theta \propto 1/m_s$ and therefore modify the magnitude of the predicted polarization splittings without altering the characteristic
charge-odd structure. The present choice provides a conservative estimate of the expected effect compared with calculations employing the much smaller current strange-quark mass. Here, $\tau_0$ and $\tau_f$ are the initial and freeze-out times of the system, respectively. $\Theta$ denotes the magnitude of the accumulated precession phase. The physical rotation angle is opposite for strange and antistrange quarks, corresponding to the transformation $\Theta\rightarrow-\Theta$. 

To estimate the sensitivity of our results to the poorly constrained magnetic-field evolution, we consider three representative magnetic-field evolution scenarios,
\begin{equation}
eB(\tau)=
\begin{cases}
eB_{0}
\left[
\dfrac{1+\tau^{2}/\tau_{B}^{2}}
{1+\tau_{0}^{2}/\tau_{B}^{2}}
\right]^{-3/2},
& \text{Vacuum},\\[0.4cm]
eB_{0}\exp\!\left[-(\tau-\tau_{0})/\tau_{B}\right],
& \text{Exponential},\\[0.35cm]
\dfrac{eB_{0}}{1+\tau^{2}/\tau_{B}^{2}},
& \text{Resistive MHD},
\end{cases}
\label{eq:Bprofiles}
\end{equation}
where $eB_{0}$ is the initial magnetic field, $\tau_{0}$ denotes the initial proper time, and $\tau_{B}$ characterizes the magnetic-field lifetime. The vacuum profile represents the rapid decay of the spectator-induced field in the absence of a conducting medium~\cite{Kharzeev:2007jp,Chen:2021nxs}, the exponential profile serves as a phenomenological benchmark~\cite{Voronyuk:2011jd,Deng:2012pc}, while the resistive magnetohydrodynamics (MHD) profile~\cite{Xu:2020sui,Wei:2024lah} incorporates the prolongation of the magnetic field due to the finite electrical conductivity of the quark--gluon plasma. Together, these scenarios span the range of magnetic-field lifetimes commonly considered in heavy-ion collisions.

The solution of Eq.~\ref{eq4} corresponds to a rotation of the polarization vector in the $x$--$z$ plane,

\begin{equation}
\begin{pmatrix}
P_{x}^{f}\\
P_{z}^{f}
\end{pmatrix}
= \begin{pmatrix}
\cos\Theta & -\sin\Theta\\
\sin\Theta & \cos\Theta
\end{pmatrix}
\begin{pmatrix}
P_{x}^{0}\\
P_{z}^{0}
\end{pmatrix},
\label{eq5}
\end{equation}

where $(P_{x}^{0},P_{z}^{0})$ and $(P_{x}^{f},P_{z}^{f})$ denote the polarization components before and after magnetic-field-induced evolution, respectively.

Equation~\ref{eq5} shows that magnetic-field-induced spin precession mixes the transverse and longitudinal polarization components, so that the final polarization retains information about the accumulated magnetic-field history.

\begin{align}
P_{x}^{s,f} &= P_{x}^{0}\cos\Theta - P_{z}^{0}\sin\Theta,\nonumber\\
P_{z}^{s,f} &= P_{x}^{0}\sin\Theta + P_{z}^{0}\cos\Theta.
\label{eq6}
\end{align}

For antistrange quarks, the opposite electric charge corresponds to the replacement $\Theta\rightarrow-\Theta$, giving

\begin{align}
P_{x}^{\bar{s},f}
&=
P_{x}^{0}\cos\Theta
+
P_{z}^{0}\sin\Theta,
\nonumber\\
P_{z}^{\bar{s},f}
&=
-
P_{x}^{0}\sin\Theta
+
P_{z}^{0}\cos\Theta.
\label{eq8}
\end{align}

The resulting charge-odd polarization splittings are

\begin{align}
\Delta P_{x}
&=
P_{x}^{\bar{s},f}-P_{x}^{s,f}
=
2P_{z}^{0}\sin\Theta,
\nonumber\\
\Delta P_{z}
&=
P_{z}^{s,f}-P_{z}^{\bar{s},f}
=
2P_{x}^{0}\sin\Theta.
\label{eq9}
\end{align}

The sign convention is chosen such that both splittings are positive for positive initial polarizations. Equation~(\ref{eq9}) constitutes the central prediction of this work: magnetic spin precession generates charge-odd polarization splittings proportional to the accumulated precession phase $\sin\Theta$. These splittings vanish in charge-averaged samples but survive in charge-resolved measurements.

An important consequence of Eq.~\ref{eq9} is the parameter-independent correlation

\begin{equation}
\frac{\Delta P_{x}}{\Delta P_{z}}
=
\frac{P_{z}^0}{P_{x}^0},
\label{eq10}
\end{equation}

obtained from the cancellation of the common precession factor $\sin\Theta$. While the individual splittings depend on the accumulated phase $\Theta$, their ratio is independent of the magnetic-field strength and time evolution. Since medium-induced spin relaxation or decoherence suppresses both $\Delta P_{x}$ and $\Delta P_{z}$ similarly, the ratio remains largely unaffected. Observation of Eq.~\ref{eq10} would therefore provide a robust experimental test of coherent magnetic spin precession in the quark--gluon plasma.\\

The magnetic field rotates an existing polarization rather than generating it. Since the field is strongest during the deconfined stage, we consider spin evolution at the quark level and assume that strange-quark polarization is largely retained during hyperon formation~\cite{Liang:2004ph,Fang:2016vpj,Becattini:2020ngo}. Existing estimates place strange-quark spin-relaxation times ($10^2$--$10^3$ fm/$c$) well above the QGP lifetime~\cite{Kapusta:2019sad}, supporting approximately coherent spin evolution. Medium-induced decoherence is therefore expected mainly to suppress the signal magnitude without altering its characteristic charge-odd structure.

\section{Results and Discussion}

\begin{figure*}
    \centering
    \includegraphics[width=0.95\textwidth]{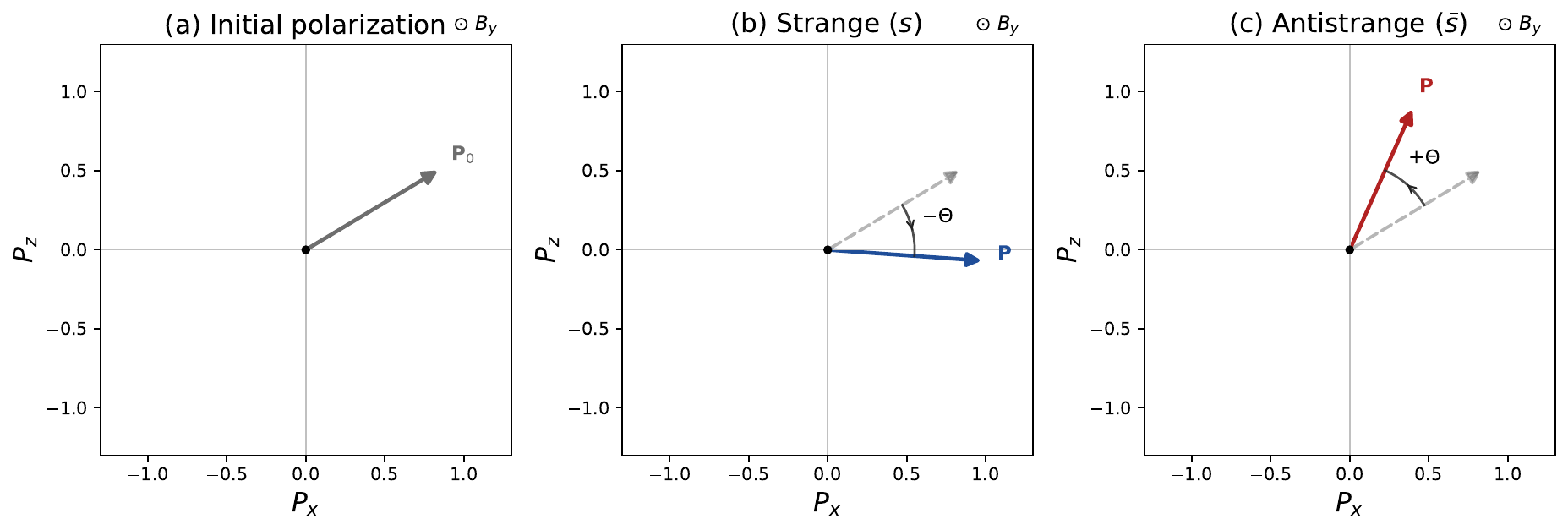}
    \caption{Representation of Larmor precession in the $x$-$z$ plane due to a magnetic field $B_y$. Panel (a) shows the initial polarization $\mathbf{P}_0$ (dashed gray). Panels (b) and (c) show rotation by $-\Theta$ (strange $s$, blue) and $+\Theta$ (antistrange $\bar{s}$, red), respectively.}
    \label{fig1}
\end{figure*}

The precession dynamics are illustrated in Fig.~\ref{fig1}. Starting from an initial polarization $\mathbf{P}_0 = (P_{x}^0, P_{z}^0)$, the magnetic field $B_y$ (out of page) rotates the polarization vector in the $x$-$z$ plane while leaving $P_{y}$ unchanged. Strange quarks precess clockwise by an angle $-\Theta$, while antistrange quarks precess counterclockwise by $+\Theta$. This opposite rotation generates charge-odd splittings, each proportional to $\sin\Theta$.

\begin{figure}[ht!]
    \centering
    \includegraphics[width=0.4\textwidth]{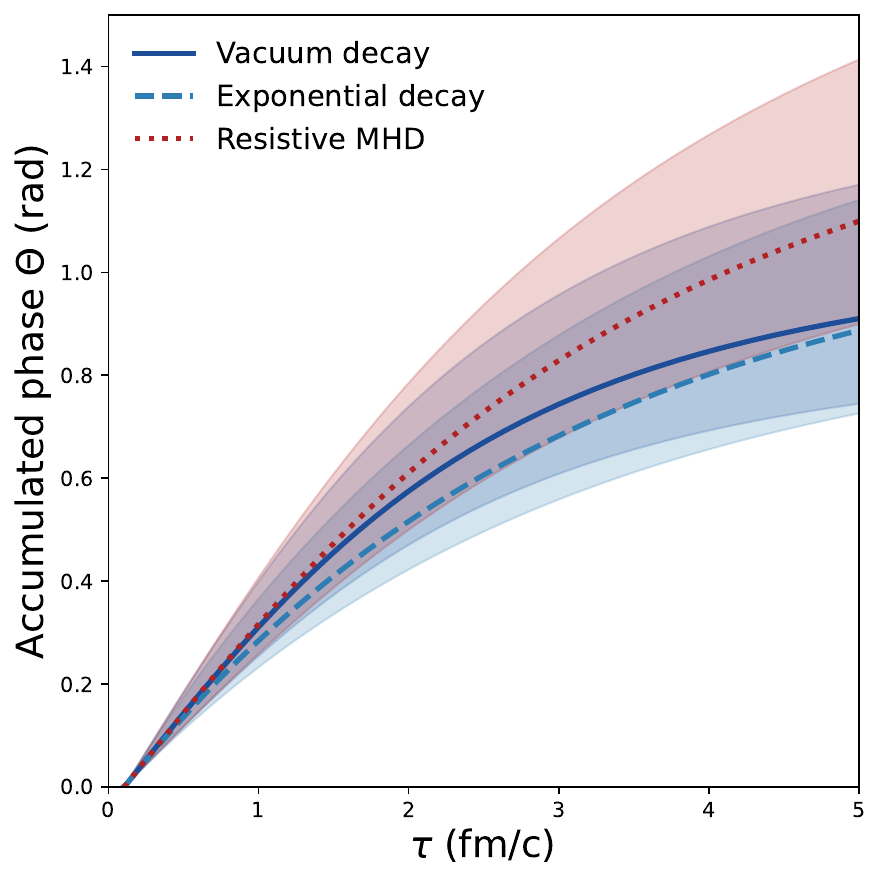}
    \caption{Accumulated Larmor precession phase $\Theta(\tau)$ as a function of proper time for three representative magnetic-field evolution scenarios: vacuum decay (solid), exponential decay (dashed), and resistive MHD (dotted).}
    \label{fig2}
\end{figure}

\begin{figure*}[ht!]
    \centering
    \includegraphics[width=0.9\textwidth]{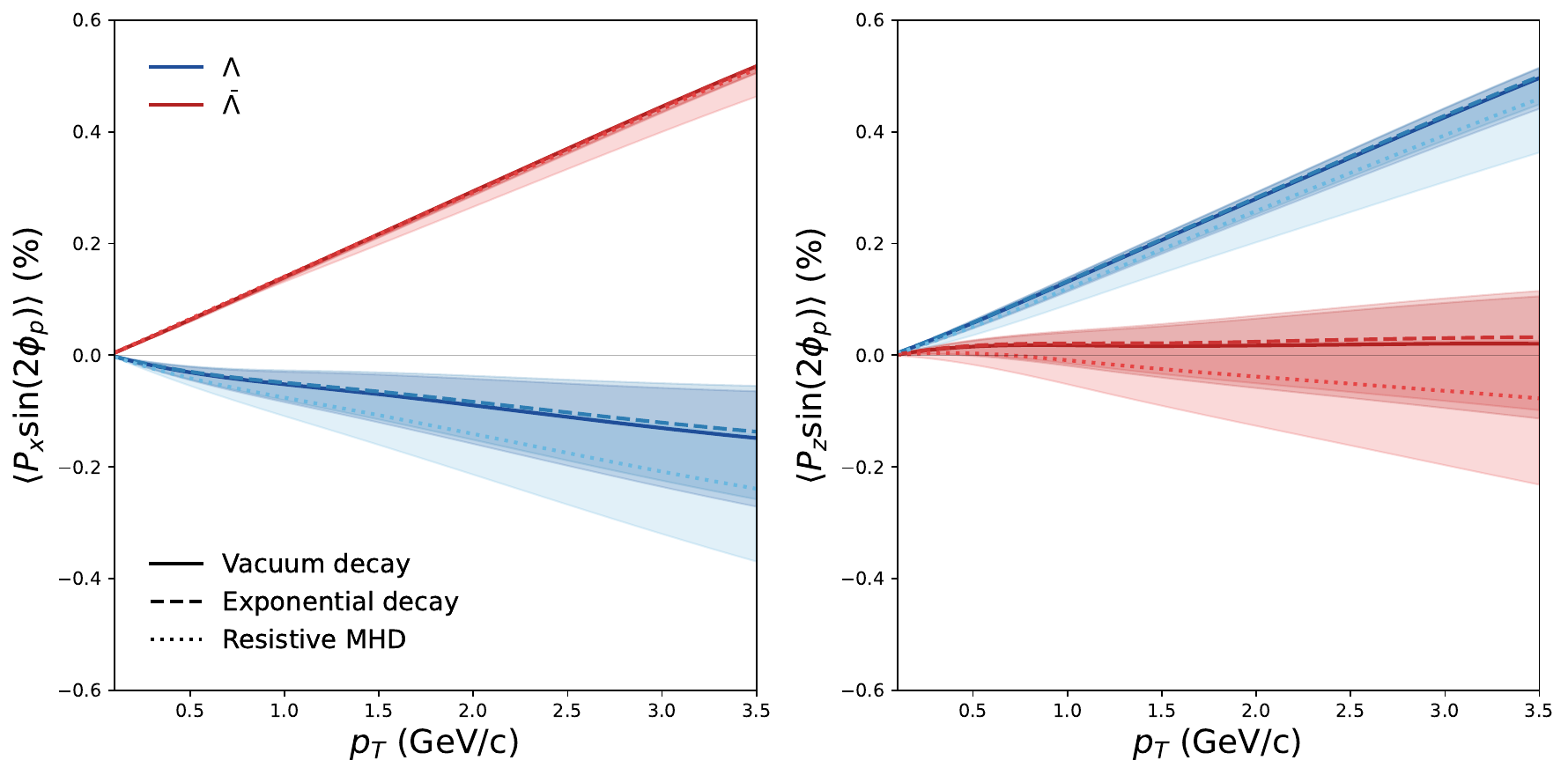}
    \caption{Polarization mixing of $\Lambda$ (blue) and $\bar{\Lambda}$ (orange) vs $p_{\rm T}$ for three magnetic field decay scenarios: vacuum (solid line), exponential (dashed line), and resistive MHD (dotted line). Left and right panels show $P_{x}$ and $P_{z}$ components, respectively. The uncertainty bands are for the strange quark effective mass range from 0.35 GeV to 0.55 GeV.}
    \label{fig3}
\end{figure*}

The three magnetic-field evolution scenarios considered here span the range of behaviors commonly discussed in the literature. Because the time evolution of the magnetic field in heavy-ion collisions remains uncertain, substantial uncertainty persists in the accumulated precession phase $\Theta$. Consequently, the magnitude of the predicted polarization splitting should be regarded as model dependent. This motivates the search for observables that are insensitive to the details of the magnetic-field evolution, leading naturally to the parameter-independent correlation discussed in Eq.~\ref{eq10}. Event-by-event fluctuations of the electromagnetic field provide an additional source of uncertainty that is not included in the present treatment. Such fluctuations may modify the accumulated precession phase on an event-by-event basis and therefore reduce the average magnitude of the polarization splitting after event averaging. However, the characteristic charge-odd structure of Eqs.~\ref{eq9} and \ref{eq10} originates from the opposite electric charges of strange and antistrange quarks and is therefore expected to remain robust against fluctuations that preserve the dominant out-of-plane orientation of the magnetic field.

Fig.~\ref{fig2} shows the accumulated precession phase $\Theta(\tau)$ for different field profiles. Although all profiles are normalized to the same initial field strength, they generate significantly different integrated phases. Since the polarization rotation is governed by $\Theta \propto \int d\tau\,B(\tau)$, the final-state polarization retains memory of the full electromagnetic history of the medium. The magnitude of the resulting charge-odd polarization splitting is controlled by $\Delta P \sim 2P_0\sin\Theta$, where $P_0$ denotes the initial polarization. Consequently, even moderate accumulated phases can generate sizable particle--antiparticle differences, making the effect potentially accessible to experiment.

To provide a realistic baseline for the initial polarization, we employ hydrodynamic input from Ref.~\cite{Arslan:2025tan}, which reproduces STAR measurements of global and longitudinal polarization at $\sqrt{s_{\rm NN}}=200$~GeV. The initial polarization vector $(P_{x}^0,P_{z}^0)$ is evolved through the Larmor precession mechanism to obtain the final-state polarization of $\Lambda$ and $\bar{\Lambda}$ hyperons. Figure~\ref{fig3} presents the resulting polarization patterns. The constituent strange and antistrange quarks undergo opposite precession phases, which are transferred to the final $\Lambda$ and $\bar{\Lambda}$ polarizations. This leads to a charge-dependent redistribution of polarization between transverse and longitudinal components. The resulting splitting grows with the accumulated phase and produces correlated modifications in both $P_{x}$ and $P_{z}$, reflecting the underlying spin-precession dynamics. At $p_{\rm T} = 3.5$ GeV/$c$, all the magnetic field scenarios yields substantial splittings in $\Lambda$ and $\bar{\Lambda}$ polarizations in the range of $0.5\%$, well within experimental reach. The predicted charge-odd polarization splittings are of order $10^{-3}$--$10^{-2}$, comparable to the magnitude of the longitudinal polarization signals already measured by STAR and ALICE. This suggests that the effect may be accessible in future high-statistics charge-resolved hyperon polarization measurements.

The numerical magnitude of the predicted splittings also depends on the choice of the effective strange-quark mass. Varying $m_s$ within a reasonable quasiparticle range of approximately $0.3$--$0.5$ GeV changes the accumulated
precession phase and therefore the absolute values of $\Delta P_x$ and $\Delta P_z$. However, the qualitative prediction of opposite $\Lambda$ and $\bar{\Lambda}$ spin rotations, as well as the parameter-independent correlation remains unchanged.

A key feature of the proposed mechanism is a coherent transfer of polarization between transverse and longitudinal components driven by Larmor precession, which is absent in conventional vorticity-induced spin-generation scenarios at leading order. Although spin relaxation and decoherence could, in principle, attenuate the signal, kinetic theory estimates for strange-quark helicity-flip and spin-relaxation processes in a rotating QGP yield equilibration times of order $10^2$--$10^3$~fm/$c$, well above the typical QGP lifetime of $\sim 10$~fm/$c$~\cite{Kapusta:2019sad}. Within this hierarchy, spin evolution remains effectively coherent over the timescales relevant for magnetic-field-induced precession, with medium effects entering only as subleading damping corrections. In experimental measurements, the observed $\Lambda$ sample contains feed-down contributions from heavier strange hadrons. Such effects may modify the quantitative magnitude of the predicted polarization splitting and should be included in future phenomenological studies. Nevertheless, the charge-odd polarization pattern generated by opposite precession phases of strange and antistrange quarks is expected to remain a generic feature of the mechanism proposed here.

The proposed observable is particularly relevant for the high-statistics heavy-ion data collected during the LHC Run~3 program. Because the charge-odd contribution cancels in the combined $\Lambda+\bar{\Lambda}$ sample, a dedicated separation of $\Lambda$ and $\bar{\Lambda}$ is required. Future measurements of the correlated splittings $\Delta P_{x}$ and $\Delta P_{z}$ would provide a test of Eq.~\ref{eq10}. Direct coupling of the hyperon magnetic moment to the external magnetic field may also generate particle--antiparticle differences. However, such contributions are expected to be aligned primarily with the local magnetic-field direction. In contrast, the mechanism proposed here is driven by Larmor precession in the dominant out-of-plane field $B_y$, which dynamically transfers polarization between different spatial components and generates correlated charge-odd splittings in both $P_{x}$ and $P_{z}$. The observation of such correlated splittings would therefore provide a distinctive signature of magnetic-field-induced spin precession in the quark--gluon plasma.

\section{Conclusion and Outlook}

We have shown that Larmor precession of polarized strange quarks in the magnetic field generated in relativistic heavy-ion collisions produces charge-odd polarization splittings between $\Lambda$ and $\bar{\Lambda}$ hyperons through the mixing of transverse and longitudinal polarization components. The proposed mechanism implies that hyperon polarization retains information about the time-integrated magnetic field experienced during the evolution of the medium, rather than solely reflecting the initial polarization generated in the collision. The mechanism further predicts a parameter-independent correlation between the polarization splittings $\Delta P_{x}$ and $\Delta P_{z}$, providing a robust experimental consistency test of coherent magnetic spin precession in the quark--gluon plasma. Charge-resolved measurements of $\Lambda$ and $\bar{\Lambda}$ polarization therefore provide a new probe of magnetic-field-induced spin dynamics and the space--time evolution of magnetic fields in relativistic heavy-ion collisions.

Future studies incorporating event-by-event magnetic-field fluctuations, spin relaxation, hadronization, and realistic hydrodynamic evolution will enable quantitative comparisons with experimental data. Conversely, the absence of the predicted charge-odd polarization splittings would imply either a small accumulated precession phase ($\Theta \ll 1$) or substantial spin decoherence during the QGP evolution, thereby constraining both the magnetic-field lifetime and spin-transport properties of the medium.
\\

\section*{Acknowledgement}
DS acknowledges the support from the postdoctoral fellowship of the DGAPA UNAM.
\\


\begin{thebibliography}{}

%\cite{Tuchin:2013apa}
\bibitem{Tuchin:2013apa}
K.~Tuchin,
%``Time and space dependence of the electromagnetic field in relativistic heavy-ion collisions,''
Phys. Rev. C \textbf{88}, 024911 (2013).

%\cite{McLerran:2013hla}
\bibitem{McLerran:2013hla}
L.~McLerran and V.~Skokov,
%``Comments About the Electromagnetic Field in Heavy-Ion Collisions,''
Nucl. Phys. A \textbf{929}, 184 (2014).


%\cite{Basar:2012bp}
\bibitem{Basar:2012bp}
G.~Basar, D.~Kharzeev and V.~Skokov,
%``Conformal anomaly as a source of soft photons in heavy ion collisions,''
Phys. Rev. Lett. \textbf{109}, 202303 (2012)

%\cite{Inghirami:2016iru}
\bibitem{Inghirami:2016iru}
G.~Inghirami, L.~Del Zanna, A.~Beraudo, M.~H.~Moghaddam, F.~Becattini and M.~Bleicher,
%``Numerical magneto-hydrodynamics for relativistic nuclear collisions,''
Eur. Phys. J. C \textbf{76}, 659 (2016).

%\cite{Huang:2022qdn}
\bibitem{Huang:2022qdn}
A.~Huang, D.~She, S.~Shi, M.~Huang and J.~Liao,
%``Dynamical magnetic fields in heavy-ion collisions,''
Phys. Rev. C \textbf{107}, 034901 (2023).


%\cite{Sahu:2025tmb}
\bibitem{Sahu:2025tmb}
D.~Sahu,
%``Barnett effect as a new source of magnetic field in heavy-ion collisions,''
Phys. Lett. B \textbf{872}, 140081 (2026).

%\cite{Sahu:2026lbv}
\bibitem{Sahu:2026lbv}
D.~Sahu,
%``Einstein-de Haas effect and induced rotation in QCD matter,''
[arXiv:2605.03093].

%\cite{Sahu:2026ixs}
\bibitem{Sahu:2026ixs}
D.~Sahu and C.~R.~Singh,
%``Einstein-de Haas effect and induced rotation in an evolving magnetized QCD matter,''
[arXiv:2606.09760].

%\cite{STAR:2017ckg}
\bibitem{STAR:2017ckg}
L.~Adamczyk \textit{et al.} [STAR],
%``Global $\Lambda$ hyperon polarization in nuclear collisions: evidence for the most vortical fluid,''
Nature \textbf{548}, 62 (2017).

%\cite{Becattini:2013vja}
\bibitem{Becattini:2013vja}
F.~Becattini, L.~Csernai and D.~J.~Wang,
%``$\Lambda$ polarization in peripheral heavy ion collisions,''
Phys. Rev. C \textbf{88}, 034905 (2013)
[erratum: Phys. Rev. C \textbf{93}, 069901 (2016)].

%\cite{Sahoo:2024egx}
\bibitem{Sahoo:2024egx}
B.~Sahoo, C.~R.~Singh and R.~Sahoo,
%``Estimating longitudinal polarization of {\ensuremath{\Lambda}} and {\ensuremath{\Lambda}}{\textasciimacron} hyperons at relativistic energies using hydrodynamic and transport models,''
Phys. Scripta \textbf{100}, 065310 (2025).

%\cite{Sahoo:2024yud}
\bibitem{Sahoo:2024yud}
B.~Sahoo, C.~R.~Singh and R.~Sahoo,
%``Impact of strong magnetic field, baryon chemical potential, and medium anisotropy on polarization and spin alignment of hadrons,''
Eur. Phys. J. C \textbf{85}, 580 (2025).

%\cite{STAR:2018gyt}
\bibitem{STAR:2018gyt}
J.~Adam \textit{et al.} [STAR],
%``Global polarization of $\Lambda$ hyperons in Au+Au collisions at $\sqrt{s_{_{NN}}}$ = 200 GeV,''
Phys. Rev. C \textbf{98}, 014910 (2018).



%\cite{Becattini:2017gcx}
\bibitem{Becattini:2017gcx}
F.~Becattini and I.~Karpenko,
%``Collective Longitudinal Polarization in Relativistic Heavy-Ion Collisions at Very High Energy,''
Phys. Rev. Lett. \textbf{120}, 012302 (2018).

%\cite{Florkowski:2017ruc}
\bibitem{Florkowski:2017ruc}
W.~Florkowski, B.~Friman, A.~Jaiswal and E.~Speranza,
%``Relativistic fluid dynamics with spin,''
Phys. Rev. C \textbf{97}, 041901 (2018).

%\cite{Weickgenannt:2019dks}
\bibitem{Weickgenannt:2019dks}
N.~Weickgenannt, X.~L.~Sheng, E.~Speranza, Q.~Wang and D.~H.~Rischke,
%``Kinetic theory for massive spin-1/2 particles from the Wigner-function formalism,''
Phys. Rev. D \textbf{100}, 056018 (2019).

%\cite{Liu:2021uhn}
\bibitem{Liu:2021uhn}
S.~Y.~F.~Liu and Y.~Yin,
%``Spin polarization induced by the hydrodynamic gradients,''
JHEP \textbf{07}, 188 (2021).

%\cite{Liang:2004ph}
\bibitem{Liang:2004ph}
Z.~T.~Liang and X.~N.~Wang,
%``Globally polarized quark-gluon plasma in non-central A+A collisions,''
Phys. Rev. Lett. \textbf{94}, 102301 (2005)
[erratum: Phys. Rev. Lett. \textbf{96}, 039901 (2006)].

%\cite{Fang:2016vpj}
\bibitem{Fang:2016vpj}
R.~h.~Fang, L.~g.~Pang, Q.~Wang and X.~n.~Wang,
%``Polarization of massive fermions in a vortical fluid,''
Phys. Rev. C \textbf{94}, 024904 (2016).

%\cite{Bhadury:2022ulr}
\bibitem{Bhadury:2022ulr}
S.~Bhadury, W.~Florkowski, A.~Jaiswal, A.~Kumar and R.~Ryblewski,
%``Relativistic Spin Magnetohydrodynamics,''
Phys. Rev. Lett. \textbf{129}, 192301 (2022).


\bibitem{Jackson} 
J. D. Jackson, \textit{Classical Electrodynamics}, 3rd ed. (Wiley, New York, 1998).

%\cite{ALICE:2021pzu}
\bibitem{ALICE:2021pzu}
S.~Acharya \textit{et al.} [ALICE],
%``Polarization of $\Lambda$ and $\bar \Lambda$ Hyperons along the Beam Direction in Pb-Pb Collisions at $\sqrt {s_{NN}}$=5.02{\,}{\,}TeV,''
Phys. Rev. Lett. \textbf{128}, 172005 (2022).


%\cite{STAR:2019erd}
\bibitem{STAR:2019erd}
J.~Adam \textit{et al.} [STAR],
%``Polarization of $\Lambda$ ($\bar{\Lambda}$) hyperons along the beam direction in Au+Au collisions at $\sqrt{s_{_{NN}}}$ = 200 GeV,''
Phys. Rev. Lett. \textbf{123}, 132301 (2019).

%\cite{Kharzeev:2007jp}
\bibitem{Kharzeev:2007jp}
D.~E.~Kharzeev, L.~D.~McLerran and H.~J.~Warringa,
%``The Effects of topological charge change in heavy ion collisions: 'Event by event P and CP violation',''
Nucl. Phys. A \textbf{803}, 227 (2008).

%\cite{Hattori:2016emy}
\bibitem{Hattori:2016emy}
K.~Hattori and X.~G.~Huang,
%``Novel quantum phenomena induced by strong magnetic fields in heavy-ion collisions,''
Nucl. Sci. Tech. \textbf{28}, 26 (2017).

%\cite{Becattini:2022zvf}
\bibitem{Becattini:2022zvf}
F.~Becattini,
%``Spin and polarization: a new direction in relativistic heavy ion physics,''
Rept. Prog. Phys. \textbf{85}, 122301 (2022).

%\cite{Becattini:2020ngo}
\bibitem{Becattini:2020ngo}
F.~Becattini and M.~A.~Lisa,
%``Polarization and Vorticity in the Quark{\textendash}Gluon Plasma,''
Ann. Rev. Nucl. Part. Sci. \textbf{70}, 395 (2020).



%\cite{Bargmann:1959gz}
\bibitem{Bargmann:1959gz}
V.~Bargmann, L.~Michel and V.~L.~Telegdi,
%``Precession of the polarization of particles moving in a homogeneous electromagnetic field,''
Phys. Rev. Lett. \textbf{2}, 435 (1959).

%\cite{Peshier:1995ty}
\bibitem{Peshier:1995ty}
A.~Peshier, B.~Kampfer, O.~P.~Pavlenko and G.~Soff,
%``A Massive quasiparticle model of the SU(3) gluon plasma,''
Phys. Rev. D \textbf{54}, 2399 (1996).

%\cite{Bluhm:2007cp}
\bibitem{Bluhm:2007cp}
M.~Bluhm and B.~Kampfer,
%``Quasiparticle model of quark-gluon plasma at imaginary chemical potential,''



%\cite{Chen:2021nxs}
\bibitem{Chen:2021nxs}
Y.~Chen, X.~L.~Sheng and G.~L.~Ma,
%``Electromagnetic fields from the extended Kharzeev-McLerran-Warringa model in relativistic heavy-ion collisions,''
Nucl. Phys. A \textbf{1011}, 122199 (2021)


%\cite{Voronyuk:2011jd}
\bibitem{Voronyuk:2011jd}
V.~Voronyuk, V.~D.~Toneev, W.~Cassing, E.~L.~Bratkovskaya, V.~P.~Konchakovski and S.~A.~Voloshin,
%``(Electro-)Magnetic field evolution in relativistic heavy-ion collisions,''
Phys. Rev. C \textbf{83}, 054911 (2011).

%\cite{Deng:2012pc}
\bibitem{Deng:2012pc}
W.~T.~Deng and X.~G.~Huang,
%``Event-by-event generation of electromagnetic fields in heavy-ion collisions,''
Phys. Rev. C \textbf{85}, 044907 (2012).


%\cite{Xu:2020sui}
\bibitem{Xu:2020sui}
K.~Xu, S.~Shi, H.~Zhang, D.~Hou, J.~Liao and M.~Huang,
%``Extracting the magnitude of magnetic field at freeze-out in heavy-ion collisions,''
Phys. Lett. B \textbf{809}, 135706 (2020).

%\cite{Wei:2024lah}
\bibitem{Wei:2024lah}
M.~Wei and L.~Yan,
%``Weak magnetic field effect on dilepton polarization in heavy-ion collisions,''
Phys. Rev. D \textbf{110}, 054024 (2024).

%\cite{Kapusta:2019sad}
\bibitem{Kapusta:2019sad}
J.~I.~Kapusta, E.~Rrapaj and S.~Rudaz,
%``Relaxation Time for Strange Quark Spin in Rotating Quark-Gluon Plasma,''
Phys. Rev. C \textbf{101}, 024907 (2020).

%\cite{Arslan:2025tan}
\bibitem{Arslan:2025tan}
A.~Arslan, W.~B.~Dong, C.~Gale, S.~Jeon, Q.~Wang and X.~Y.~Wu,
%``In-plane transverse polarization in heavy-ion collisions,''
Phys. Rev. C \textbf{113}, 034920 (2026).



\end{thebibliography}
\end{document}